\documentclass[journal]{IEEEtran}
\usepackage{multirow}
\usepackage{graphicx} % for arXiv
\usepackage{epsfig} % for postscript graphics files
\usepackage{mathptmx} % assumes new font selection scheme installed
\usepackage{times} % assumes new font selection scheme installed
\usepackage{amsmath} % assumes amsmath package installed
\usepackage{amssymb}  % assumes amsmath package installed
\usepackage{subfigure}
\usepackage{textcomp,gensymb}
\usepackage{url}
\usepackage{xcolor}
\ifCLASSINFOpdf
\else
\fi
\begin{document}
\title{Comparison of Deep Reinforcement Learning and Model Predictive Control for Adaptive Cruise Control}

\author{Yuan~Lin,~\IEEEmembership{Member,~IEEE,}, John~McPhee,
        and~Nasser~L.~Azad,~\IEEEmembership{Member,~IEEE}% <-this % stops a space
\thanks{$^{1}$Dr. Yuan Lin is a Postdoctoral Fellow in the Systems Design Engineering Department at the University of Waterloo, Ontario, Canada N2L 3G1.
        {\tt\small y428lin@uwaterloo.ca}}%
\thanks{$^{2}$Dr. John McPhee is a Professor and Canada Research Chair in the Systems Design Engineering Department at the University of Waterloo, Ontario, Canada N2L 3G1.
        {\tt\small mcphee@uwaterloo.ca}}%
\thanks{$^{3}$Dr. Nasser L. Azad is an Associate Professor in the Systems Design Engineering Department at the University of Waterloo, Ontario, Canada N2L 3G1.
        {\tt\small nlashgarianazad@uwaterloo.ca}}%
}

\markboth{}%IEEE Transactions on Intelligent Vehicles}
%\markboth{IEEE Transactions on Intelligent Transportation Systems, ~Vol.~14, No.~8, August~2015}%
{Shell \MakeLowercase{\textit{et al.}}: Bare Demo of IEEEtran.cls for IEEE Journals}



\maketitle

\begin{abstract}

This study compares Deep Reinforcement Learning (DRL) and Model Predictive Control (MPC) for Adaptive Cruise Control (ACC) design in car-following scenarios. A first-order system is used as the Control-Oriented Model (COM) to approximate the acceleration command dynamics of a vehicle. Based on the equations of the control system and the multi-objective cost function, we train a DRL policy using Deep Deterministic Policy Gradient (DDPG) and solve the MPC problem via Interior-Point Optimization (IPO). Simulation results for the episode costs show that, when there are no modeling errors and the testing inputs are within the training data range, the DRL solution is equivalent to MPC with a sufficiently long prediction horizon. Particularly, the DRL episode cost is only 5.8\% higher than the benchmark solution provided by optimizing the entire episode via IPO. The DRL control performance degrades when the testing inputs are outside the training data range, indicating inadequate generalization. When there are modeling errors due to control delays, disturbances, and/or testing with a High-Fidelity Model (HFM) of the vehicle, the DRL-trained policy performs better with large modeling errors while having similar performance as MPC when the modeling errors are small.

\end{abstract}


% \fi
\IEEEpeerreviewmaketitle

\section{Introduction}

Reinforcement learning is a learning-based method for optimal decision making and control \cite{sutton2018reinforcement}. In reinforcement learning, an agent takes an action based on the environment state and consequently receives a reward. Reinforcement learning maximizes cumulative discounted reward by learning an optimal state-action mapping policy through trial and error. The policy is trained via Bellman's principle of optimality, which dictates that the remaining actions constitute an optimal policy with regard to the state resulting from a previous action. Deep reinforcement learning (DRL), which utilizes deep (multi-layer) neural nets as policy representations, has drawn significant attention as its trained policy surpassed the best human in playing board games \cite{silver2016mastering}. Different DRL algorithms have been proposed which include Deep Q-Networks \cite{mnih2015human}, Trust Region Policy Optimization \cite{schulman2015trust}, Proximal Policy Optimization \cite{schulman2017proximal}, and Deep Deterministic Policy Gradient (DDPG) \cite{lillicrap2015continuous}. In this work, we use DDPG, which outputs continuous control actions by training a deterministic policy offline. DDPG is a popular choice for optimal control, especially for a stable dynamic system \cite{henderson2018deep}.

Model Predictive Control (MPC) represents the state of the art for the practice of real-time optimal control \cite{diehl2009efficient}. MPC benefits from a sufficiently accurate model of the plant dynamics. At each time step, a constrained optimization problem is formulated based on the plant model to minimize a defined cost function in a predictive time horizon. The optimization problem is solved online and only the first value of the solved control sequence is applied. At the next time step, this predictive control procedure is repeated with updated states. There are various methods to formulate the optimization problem with the state-space equations and the cost function, which include direct single shooting, direct multiple shooting, and direct collocation \cite{maitland2018nonlinear}. There are also various online optimization solvers for MPC, which include sequential quadratic programming and IPO \cite{diehl2009efficient}. In this work, we use IPO with direct single shooting, which solves the formulated optimization problem via Newton-Raphson's method by successively approximating the root of the cost function derivative \cite{maitland2017improving}. The IPO solution is on the interior of the set described by the inequality constraints and close to the true optimal solution.

Since both DRL and MPC can provide optimal control solutions, it is of research interest to understand their advantages and disadvantages. For our comparison, we consider solving an optimal control problem for a dynamic system represented by a system of state-space equations. We do not consider training an end-to-end (such as image-to-control-action) solution using DRL \cite{mnih2015human}. Before using an example for comparison, one could understand some known differences between the two. Firstly, MPC demands online optimization that requires relatively powerful computing devices for real-time applications, which raises monetary concerns. For automotive engineering, hardware-in-the-loop simulations are needed to verify the real-time readiness of MPC before real-world deployment \cite{vajedi2015ecological}. On the other hand, offline-trained DRL solutions are neural nets that result in very little computation time during deployment. Secondly, MPC is model-based while, up to date, DRL control solutions are black-box neural nets that lack theoretical assurance \cite{lee2019w}. In this work, we do not focus on these known differences about the computing requirements and theoretical assurance for DRL and MPC.

In this work, we focus on the optimality level (minimum episode cost) that DRL and MPC can achieve without and with modeling errors. For fair comparison, we use the same COM of the vehicle for DRL to train a policy and for MPC optimization. Most of the parameter settings are the same for both DRL and MPC except that the DRL reward utilizes a discount factor that is absent in the MPC optimization. This is due to the fact that DRL usually requires a discount factor less than one for convergence \cite{watkins1992q} while MPC normally does not include the discount factor.

We raised a few questions that guided our research: (1) When there are no modeling errors, for example, testing on the vehicle COM, is DRL or MPC better in achieving the minimum cost? We use IPO to optimize for the entire simulation episode once to obtain a benchmark solution, called the IPO solution, for both DRL and MPC. Note that the IPO solution is not a receding-horizon one since it's obtained by setting the predictive time horizon as the episode length and the optimization is solved only once. MPC usually obtains better optimality levels with longer prediction horizons. It may be interesting to see the difference between the DRL solution and MPC with different prediction horizons. The comparison of the DRL, MPC, and IPO solutions could provide insights on training policies via Bellman's principle of optimality versus optimizing via Newton-Raphon's method. It would also show the effect of the discount factor on the optimality-seeking of DRL. Additionally, we also want to investigate if the machine learning generalization issue persists in the DRL-trained neural net. When the testing inputs are outside the range of training data, the DRL control performance may be compromised and lose competitiveness to MPC.

(2) When there are modeling errors, does DRL or MPC achieve a lower cost? Modeling errors in this paper refer to the differences from the ACC car-following state-space equations. Such modeling errors include neglected control delays, disturbances, and/or the difference between the COM and HFM. In our previous work, we showed that modeling errors due to neglecting vehicle dynamics could cause significantly degraded DRL control performance \cite{lin2019longitudinal}. As both DRL and MPC suffer from performance degradation due to modeling errors \cite{sakhdari2018adaptive,lopez2019dynamic}, this work could show whether DRL or MPC is better at handling modeling errors given that most conditions are the same. It's worth mentioning that DRL has been shown to perform better than a rule-based method for lane-change control in the presence of environment noise \cite{alizadeh2019automated}.

\begin{figure}[!htbp]
\centering
\includegraphics[width=3.2in]{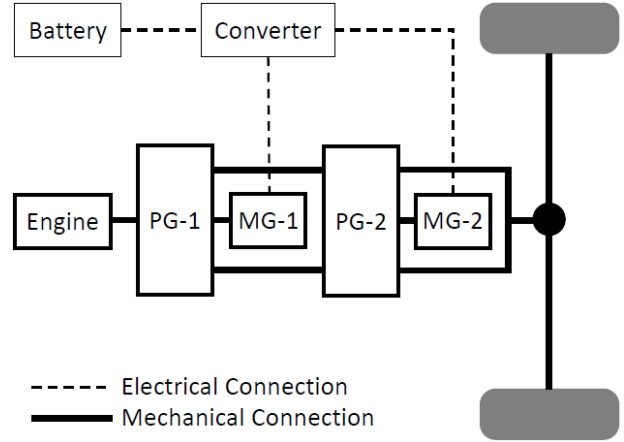}
\caption{Schematic of the HFM powertrain of a 2015 Toyota Prius power-split PHEV. In the plot, PG means planetary gear and MG means motor-generator.}
\label{fig:powertrain}
\end{figure}

To answer the raised questions, we develop both DRL and MPC controllers for ACC car-following control. Car following is one of the most common behaviors of road vehicles \cite{tajeddin2019ecological}. ACC is a type of Advanced Driver Assistance System that enables intelligent and automated driving \cite{eskandarian2012}. Automated vehicle development has been a popular interest in academia and industry as it could potentially revolutionize transportation. We develop ACC controllers for a power-split plug-in hybrid electric vehicle (PHEV), a 2015 Toyota Prius, since we have previously developed a HFM of it in MATLAB/Simulink \cite{vajedi2015ecological,taghavipour2014high}. The HFM includes control input execution delay (control delay) of 0.2s, powertrain modeling, and external resistances including aerodynamic drag and rolling resistance. Road grade is not considered as we assume flat surfaces. The complexity of the HFM can be shown by its powertrain modeling, see Fig.~\ref{fig:powertrain}. The powertrain modeling of the HFM includes the modeling of its battery, battery converter, electric motors, combustion engine, and planetary gears. In addition, the HFM includes a rule-based energy management system (charge-depletion-charge-sustaining) to determine the power demands for the battery and engine \cite{kim2011comparison}. The HFM is based on Autonomie, a MATLAB/Simulink simulation tool for automotive control developed by the Argonne National Lab. Note that the first-order vehicle COM considered in this work does not include the control delay.

We acknowledge that our comparison of MPC and DRL is limited to a certain scope when considering the effect of modeling errors. On one hand, there are more advanced MPC and DRL methodologies. Our adopted MPC methodology that includes direct single shooting and IPO is typical yet simple. Research advances in tube-based and stochastic MPC could make MPC more robust and disturbance-tolerant \cite{sakhdari2018adaptive,lopez2019dynamic,moser2017flexible}. Regarding DRL, our adopted DDPG algorithm is a cornerstone but could be polished. The use of transfer learning and/or meta-learning on the DRL-trained policy could make it better in handling modeling errors and uncertainties \cite{van2019sim,finn2017model}. On the other hand, the ACC car-following control example is a low-dimensional task with only three state variables while DRL is known to handle well higher-dimensional tasks with complex cost functions \cite{lillicrap2015continuous,lowrey2018plan}. For the scope of this work, we only consider the low-dimensional task without considering robust and stochastic MPC or transfer and meta-learning.

The main contribution of this work is the quantitative and comprehensive comparison of the well-known DRL algorithm, DDPG, and an MPC that is based on the popular IPO method. We consider the effect of the MPC prediction horizon, the generalization issue of DRL, the case of no modeling errors, and the cases of modeling errors that include the control delay, disturbances, and testing with the HFM. To our best knowledge, there is no such comparison existing in the literature. We hope that such a comprehensive comparison will serve as a useful reference for researchers working on optimal control.

Regarding the paper organization, a literature review is documented in Section II; the ACC problem formulation is in Section III; methodologies of DRL and MPC are documented in Section IV; results are shown in Section V; and Section VI draws conclusions.

% wierd non-linear system but non linear cost

\section{Literature Review}

There are only a limited number of papers in the literature that compare reinforcement learning and MPC performances. In \cite{ernst2008reinforcement}, the authors compared reinforcement learning and MPC in controlling non-linear electrical power oscillation damping. With a random tree as the policy, the reinforcement learning is not DRL. With a low-dimensional deterministic model of the system, the authors considered no modeling errors. The results show that with different parameter settings, the reinforcement learning solutions could be worse or better than MPCs with regard to the cumulative discounted cost. The authors also showed data that indicates that reinforcement learning is at least 10 times faster than MPC during testing.

In \cite{julian2019distributed}, the authors compared DRL and receding-horizon control (same as MPC) in controlling a team of unmanned aerial vehicles to maximize wild fire coverage. The authors used a stochastic model of wild fire propagation that adds randomness (disturbances) to the control. The DRL environment state is high-dimensional since it includes both images and continuous states, indicating a hybrid-input DRL control. The results show that DRL outperformed receding-horizon control by a moderate margin regarding cumulative reward.

In \cite{lowrey2018plan}, the authors compared integrated MPC-DRL and pure MPC controllers for control of high-dimensional tasks such as 3D humanoid standing up from the ground and in-hand manipulation by a five-fingered robotic hand. The integrated MPC-DRL controller is essentially a MPC controller wherein the MPC terminal cost is learned via DRL. The training and testing were based on an accurate model without considering modeling errors. The authors found that the integrated MPC-DRL controller achieved higher rewards than a pure MPC controller by a moderate margin.%used MPPI in the paper

In \cite{saxena2019driving}, the authors compared DRL and MPC for merging into dense traffic. The DRL and MPC methods do not share the same cost function. Specifically, DRL has a complex cost function including absolute-value and linear costs while MPC has a quadratic cost function. Thus, the authors did not compare the episode costs of DRL and MPC. However, the authors found that the DRL-trained policy significantly out-performed MPC regarding the rate of merging success.

In summary, there is a lack of literature on comparing DRL and MPC in a fair manner, especially in the presence of modeling errors. Our motive originates from solving a traditional optimal control problem that can be represented by state-space equations. In our work, most conditions are set to be the same for DRL and MPC for fair comparison. The HFM of Prius enables us to study the effect of practically-existing modeling errors on the control performances of DRL and MPC. These characteristics make our work different from the existing literature.

There is also limited literature on ACC car-following control using DRL. In \cite{desjardins2011cooperative}, the authors used a single-layer (non-deep) neural net as the reinforcement learning policy representation to train an ACC controller. In \cite{zhao2013supervised,zhu2018human}, naturalistic driving data was used to train human-like car-following policies using DRL. In our previous work, we trained an ACC optimal control policy with a state-space car-following model using DRL for the first time \cite{lin2019longitudinal}. Our previous work is the base for the DRL controller development in this paper. However, the car-following model in this paper considers a constant time headway instead of a constant distance headway in the previous work. The constant time headway enables the vehicle to proportionally adjust the desired inter-vehicular distance based on its speed, which is more appropriate in real-world driving.

There is a large body of literature on ACC using MPC \cite{vajedi2015ecological,sakhdari2018adaptive,luo2010model,li2010model}. In such research papers, the model-predictive ACC systems were designed with multi-objective cost functions to minimize the tracking error, energy consumption, vehicle jerk, and etc. A first-order system was usually considered to be sufficient to approximate the acceleration command dynamics of the vehicle \cite{li2017dynamical,ploeg2011design}. The first-order approximation is due to the imperfect estimation of vehicle parameters, lower-level control of acceleration and brake pedals' positions, unmodeled powertrain dynamics, and external disturbances \cite{li2010model}. Our ACC problem formulation described in the following section is similar to that from the model-predictive ACC papers.

\section{ACC Problem Formulation}

\begin{figure}[!htbp]
\centering
\includegraphics[width=3.2in]{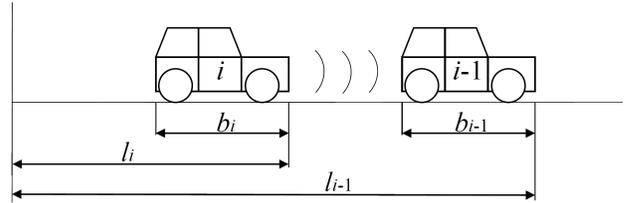}
\caption{Schematic for two-car following.}
\label{fig:schematic}
\end{figure}

In this section, we formulate the ACC problem with state-space equations and define the multi-objective cost function. We consider controlling a following vehicle $i$ to maintain a constant time gap $t_g$=1s with its immediately preceding vehicle $i-1$, see Fig.~\ref{fig:schematic}. The gap-keeping error $e$ and the velocity difference $e_v$ between the preceding and following vehicles are:
\begin{align}
\begin{split}
e &= l_{i-1} - l_{i} - b_{i-1} - (d_0 + h v_i)\\
e_v &= v_{i-1} - v_i
\end{split}
\label{eq:e}
\end{align}
where $l_{i-1}$ and $l_{i}$ are the distances traversed by the preceding $i-1$ and following $i$ vehicles, respectively, $b_{i-1}$ is the vehicle body length of the preceding vehicle $i-1$, $d_0$ is a standstill distance for safety, and $v_{i-1}$ and $v_i$ are the velocities of the preceding $i-1$ and following $i$ vehicles, respectively. Taking the derivatives of $e$ and $e_v$ and using a first-order system to approximate the acceleration dynamics \cite{ploeg2011design}, we obtain the following state-space equations for ACC:
\begin{align}
\begin{split}
\dot{e} &= v_{i-1} - v_i - t_g a_i = e_v - t_g a_i\\
\dot{e}_v &= a_{i-1} - a_i = - a_i\\
\dot{a}_i &= \frac{u_i - a_i}{\tau}
\end{split}
\label{eq:ss}
\end{align}
where $a_i$ is the acceleration of the following vehicle, $a_{i-1}$ is the acceleration of the preceding vehicle whose value is set to zero, and $u_i$ is the control input (commanded acceleration) to the following vehicle. Setting $a_{i-1}=0$ in the state-space equations depicts the situation that the preceding vehicle is running with a constant speed. When the preceding vehicle has varying speeds, its acceleration values can be considered as disturbances. They contribute to the modeling errors and are helpful for us to compare DRL and MPC on handling modeling errors.

The third equation in the state-space equations is the first-order system approximating the powertrain dynamics, also called the longitudinal COM of the following vehicle in this paper. The first-order-system approximation is on the basis that there is a low-level controller that minimizes the error between the control input and the actual acceleration. For our HFM, we designed a low-level controller with feedback and feedforward within the HFM. The feedback portion of the low-level controller is a PI controller that minimizes the error between the control input and actual acceleration; the feedforward portion equals the external resistant forces of the HFM which include the aerodynamic drag and the rolling resistance. We test three drive cycles, the Highway Fuel Economy Test (HWFET), the Federal Test Procedure 75 (FTP-75), and the US06 Supplemental Federal Test Procedure, on the HFM to obtain data of $u_i$ and $a_i$ for system identification of $\tau$. The $\tau$ values identified are 0.12s, 0.09s, and 0.13s for the three drive cycles respectively. Thus, we choose a constant $\tau$=0.1s which is considered as the average. In \cite{ploeg2011design}, the authors found through road tests that a first-order system with $\tau$=0.1s provided a good match to the longitudinal behaviors of a Toyota Prius; they also found the control delay to be 0.2s, which is what we used in the HFM.

The ACC cost function has three objectives: minimizing gap-keeping error $e$, control effort $u_i$, and jerk which is the rate of acceleration $j_i = \dot{a}_i$. The cost function is defined as:
\begin{equation}
\begin{split}
c = &\sum \alpha \sqrt{(\frac{e_{t}}{e_{nmax}})^2 + \epsilon} + \beta \sqrt{(\frac{u_{i}}{u_{min}})^2 + \epsilon}\\
& + \gamma \sqrt{[\frac{\dot{a}_i}{(u_{max}-u_{min})/\Delta t}]^2 + \epsilon}
\end{split}
\label{eq:cost_function}
\end{equation}
where $\alpha=\beta=\gamma=1/3$ are the weights, $\epsilon=10^{-8}$ is a small value, $\Delta t$=0.1s is the time step, $e_{nmax}$=15m is the nominal maximum gap-keeping error, and $u_{max}=2$m/s$^2$ and $u_{min}=-3$m/s$^2$ are the allowed maximum and minimum control input values, respectively. Table~\ref{table:acc_para} summarizes the ACC parameter values.

\begin{table}[!htbp]
\caption{ACC parameter values.}
\begin{center}
\begin{tabular}{|c|c|}
\hline
Constant time gap $t_g$ & 1s \\
\hline
First-order-system COM time constant $\tau$ & 0.1s \\
\hline
Time step $\Delta t$ & 0.1s \\
\hline
Allowed maximum control input $u_{max}$ & 2m/s$^2$ \\
\hline
Allowed minimum control input $u_{min}$ & -3m/s$^2$ \\
\hline
Nominal maximum gap-keeping error $e_{nmax}$ & 15m \\
\hline
Weight for gap-keeping error cost $\alpha$ & 1/3 \\
\hline
Weight for control action cost $\beta$ & 1/3 \\
\hline
Weight for jerk cost $\gamma$ & 1/3 \\
\hline
\end{tabular}
\end{center}
\label{table:acc_para}
\end{table}

The value of $e_{nmax}$ is chosen such that it is larger than most possible gap-keeping errors. During training, the gap-keeping errors could be larger than the $e_{nmax}$ as the DRL agent explores an acceleration policy. This is why we call $e_{nmax}$ the nominal maximum instead of the true maximum. The sum of the weights in the cost function is 1 as we desire the cost at a time step to be within [0,1]. The three constant weights are chosen to be equal, but without a specific purpose. The weights could be adjusted if desired to balance the minimization of the error, control action, and jerk. It's worth mentioning that the weights could be dynamically tuned based on the states using another layer of DRL for improved performance of objective tracking \cite{ure2019enhancing}.

The minimization of the control effort $u_{i}$ indirectly suggests energy-efficient driving for a PHEV. We acknowledge that more sophisticated modeling of the Prius hybrid energy management system could be considered to minimize the monetary cost of both fuel and battery usage \cite{vajedi2015ecological,sakhdari2018adaptive,tajeddin2019ecological}. Since this paper is focused on the optimality seeking of DRL versus MPC without and with modeling errors, the more sophisticated monetary cost calculation may not be necessary. We think that the same conclusion would be drawn about comparing DRL and MPC in optimality seeking even if we consider the hybrid monetary cost.

The cost function is almost the same as an absolute-value cost function except for having $\epsilon$. The reason to add $\epsilon$ is to create a differentiable and smooth cost function such that IPO could converge. A pure absolute-value cost function is not differentiable nor smooth at the minimum, which prevents IPO from converging. The $\epsilon$ value is chosen to be very small to preserve proximity to an absolute-value cost but yet allows IPO to converge quickly. A quadratic cost function that is popular for MPC is not used since it causes significant steady-state errors of DRL solutions \cite{engel2014line}. We do not use different cost functions for DRL and MPC as it would be meaningless to compare their episode costs.

DRL utilizes the notion of reward $r$ which is the negative value of MPC cost $c$. That is, $r=-c$. At each time step $t=0,1,2,...,T$ where $T$ is the episode termination time step, the DRL reward $r_t$ is discounted as $\gamma^t r_t$ with the discount factor $\gamma$=0.99. The discount factor is chosen as $\gamma$=0.99 as it is suggested that 0.99 is among the largest that could lead to the highest reward with convergence stability \cite{franccois2015discount}. Recall that the discount factor is not used in MPC.

Hard constraints that demand set conditions to be met are not considered in the control problem formulation here. MPC is superior than other classical optimal control methods such as Linear-Quadratic Regulator due to its handling of hard constraints. However, up to date, hard constraints are still under research and cannot be guaranteed for DRL \cite{achiam2017constrained}. In order for reasonable comparison, hard constraints are not included for either MPC or DRL. Instead, we include the constraints in the multi-objective cost function as soft ones. For example, the cost for vehicle jerk is included in the cost function in order to reduce the jerk values.

% Does my mom understand?

\section{Methodologies}

In this section, the DRL and MPC methodologies to solve the optimal control problem of ACC are explained. Since both DRL and MPC are based on discrete time, Runge–Kutta-4 (RK4) is used to discretize the ACC state-space equations. For both DRL (training and testing) and MPC, the discrete update time step is $\Delta t=0.1s$. In other words, both DRL and MPC have the same control update frequency $1/\Delta t$=10Hz.

\subsection{DRL}

Reinforcement Learning is formulated as a Markov Decision Process: at time $t$, given the environment state $s_t$, an agent takes an action $a_t$ based on a policy $\mu$, resulting in a new state $s_{t+1}$ and a reward $r_t$. Reinforcement Learning learns a state-action mapping policy that maximizes the cumulative discounted reward $\sum_{t=0}^{t=T} \gamma^{\,t} r_t$. The specific reinforcement learning algorithm we use, DDPG, has two networks: the actor $\phi$ and critic $\theta$ networks \cite{lillicrap2015continuous}. The critic network, also called the Q-network, is trained based on Bellman's principle of optimality. Specifically, the Q-value for a state-action pair is defined as the cumulative discounted reward from time $t$: $Q_{\theta}(s_t,a_t) = \sum_{\tau = t}^{\tau = T} \gamma^{\,\tau - t} r_{\tau}$ where $\tau$ denotes a time step between $t$ and $T$. The critic network parameters $\theta$ are updated by minimizing the loss $L_t = r_t + \gamma Q_{\theta}(s_{t+1},\mu_{\phi}(s_{t+1})) - Q_{\theta}(s_t,a_t)$ using gradient descent with respect to $\theta$. The actor network, also called the policy network, is updated by taking a gradient ascent on the Q-value $Q_{\theta}(s_{t},\mu_{\phi}(s_{t}))$ with respect to $\phi$.

Several techniques are used to improve training stability and convergence. They include target networks, mini-batch gradient descent, experience replay, batch normalization, and addition of Gaussian noise to the action for exploration \cite{mnih2015human,ioffe2015batch,buechel2018deep}. Both the actor and critic neural nets have 2 hidden layers with 64 linear rectifier neurons for each layer. Table~\ref{table:ddpg_para} shows the DDPG parameter values. The DDPG hyper-parameter values are selected based on the values suggested in the original DDPG paper \cite{lillicrap2015continuous} and fine-tuned through trial and error. Note that DDPG is sensitive to its hyper-parameters. Thus, careful fine-tuning helps achieve better performance of the DRL controller. Even with fine-tuned hyper-parameters, we found that DDPG converged to slightly different sub-optimal policies each time, due to the slightly different episode costs in testing. Thus, we performed training 3 times and selected the best trained policy leading to the lowest episode cost. Additionally, before discounting, the reward is clipped to have values within [-1,0] since it is suggested that sudden large changes of reward values decrease training stability \cite{van2016learning}.

\begin{table}[!htbp]
\caption{DDPG parameter values.}
\begin{center}
\begin{tabular}{|c|c|}
\hline
Target network update coefficient & 0.001\\
\hline
Reward discount factor & 0.99\\
\hline
Actor learning rate & 0.0001\\
\hline
Critic learning rate & 0.001\\
\hline
Experience replay memory size & 500000\\
\hline
Mini-batch size & 64\\
\hline
Actor Gaussian noise mean & 0\\
\hline
Actor Gaussian noise standard deviation & 0.02\\
\hline
\end{tabular}
\end{center}
\label{table:ddpg_para}
\end{table}

\begin{figure}[!htbp]
\centering
\includegraphics[width=3.2in]{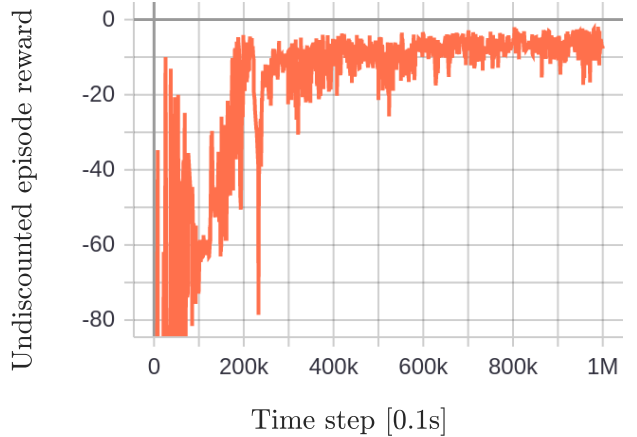}
\caption{Undiscounted episode reward during DRL training.}
\label{fig:reward}
\end{figure}

For training the ACC optimal control policy, the reinforcement learning state evolution is based on the ACC state-space equations with $a_{i-1}=0$. The DRL inputs are the values of the three states and the DRL output is the control input. For each training episode, the initial conditions for the states $e_0$, $e_{v,0}$, and $a_{i,0}$ are randomly distributed in [-5,5]m, [-5,5]m/s, and [-3,2]m/s$^2$, respectively. The training initial conditions are considered to represent normal car-following scenarios. The training episode length is 20s. We trained the DDPG policy for 1 million time steps and observed convergence of the cumulative reward. The undiscounted cumulative reward during the training is shown in Fig.~\ref{fig:reward}. The reason to show the undiscounted cumulative reward is that changes of the undiscounted cumulative reward are more observable during training. The training took 37 minutes on a desktop computer with a 16-core (32-thread) AMD processor and a Nvidia GeForce RTX GPU.

\subsection{MPC}

With the ACC state-space equations and the cost function, we formulate the optimization problem using direct single shooting for a certain prediction horizon. Then the optimization problem is solved using IPO. The first value of the solved control sequence is applied. This procedure is repeated as the receding-horizon MPC. Note that we also use IPO to solve for the optimization solution just once for the entire episode as the benchmark solution. The benchmark is denoted as IPO solution for the rest of the paper. The MPC optimization problem is formulated and solved using the open-source symbolic framework CasADi \cite{Andersson2018}. CasADi can be installed in MATLAB to provide functions of various shooting and optimization methods for programmers to solve non-linear optimization problems.

% does your mom understand?

\section{Results}

This section shows the testing results of the DRL and MPC controllers without and with modeling errors. To exclude modeling errors, the COM-trained controllers are tested on the exact ACC state-space equations; that is, the controllers are tested on the COM with $a_{i-1}=0$. Such testing allows us to better understand the inherent differences of DRL and MPC without the effect of modeling errors. We consider two cases of testing without modeling errors. For the first case, the testing initial conditions are within the range of the training ones that represent normal car following. For the second case, the testing initial conditions are outside the range of the training ones, which enables us to investigate the generalization issue of the DRL-trained neural net. For this case, we consider cut-in scenarios with large negative gap-keeping errors that are outside training data range.

To include modeling errors, we consider testing with differences from the ACC state-space equations in three controlled experiments. In the first controlled experiment, the control delay $\tau_d$ is added to the COM to investigate the impact of different control delay values on the DRL and MPC controllers' performances. In the second controlled experiment, the HFM is used to replace the COM to investigate the impact of the unmodeled vehicle characteristics such as the vehicle power limit. In the third controlled experiment, the testing is conducted with both the HFM and drive cycles. In this case, the preceding vehicle's velocity profile follows the drive cycles, meaning that $a_{i-1}\neq$0. For drive-cycle testing, the modeling errors are larger due to the non-zero $a_{i-1}$.

\subsection{Testing without modeling errors - a case study with a single initial condition}

Without modeling errors, the DRL and MPC controllers are tested on the exact ACC state-space equations. To understand the effect of the MPC prediction horizon and the time response of the states, we present a case study of a single initial condition to compare DRL and MPC control performances.

\begin{figure}[!htbp]
\centering
\includegraphics[width=3.2in]{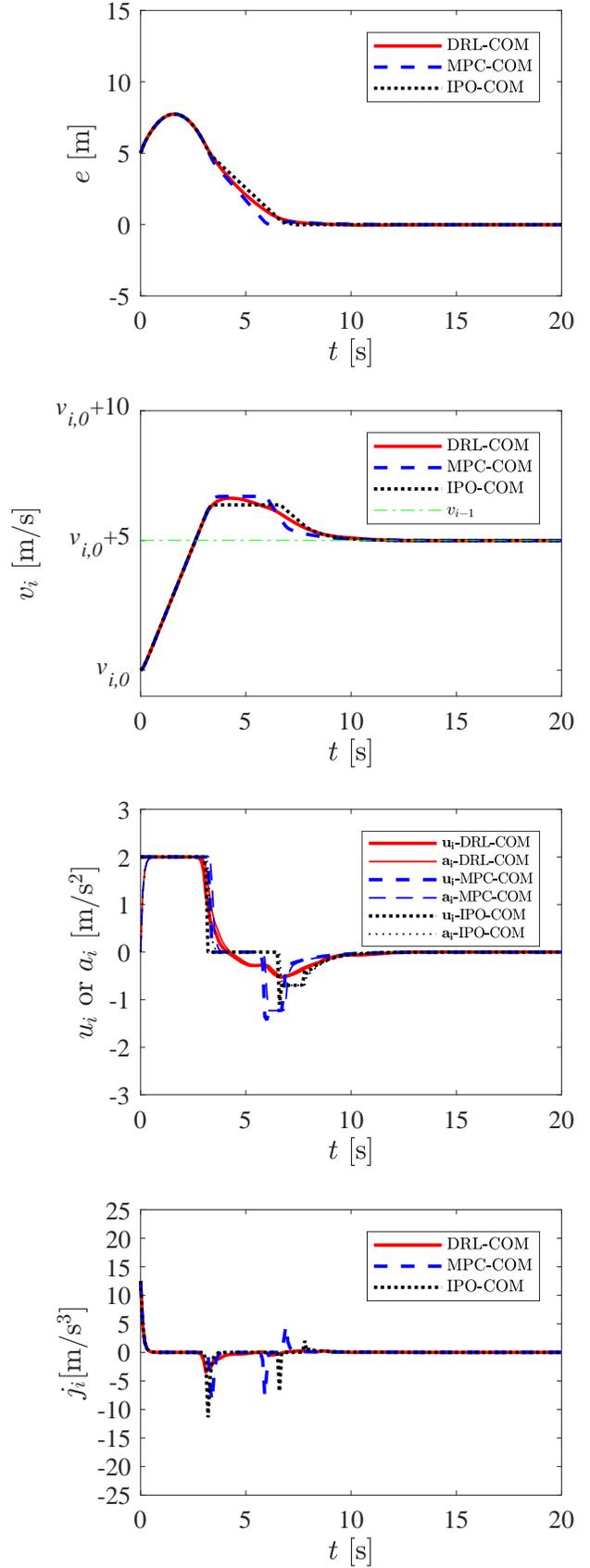}
\caption{Testing results when there are no modeling errors for the single initial condition considered. In the plots, MPC $h$=2.8s. When MPC $h$=5s, the MPC results are almost identical to those of IPO. The symbol $v_{i,0}$ denotes the initial velocity of the following vehicle $i$.}
\label{fig:com_episode}
\end{figure}

The single initial condition is $[e_0, e_{v,0}, a_{i,0}]$ = [5m,5m/s,0m/s$^2$], which is within the range of the training initial conditions. The results of the corresponding episode are plotted in Fig.~\ref{fig:com_episode}. The curves for the three methods, DRL, MPC, and IPO, look similar. In the third plot of control input and acceleration values, the DRL curve exhibits an uncommon shape, indicating that the trained policy is not model-based. The plotted MPC solution has a prediction horizon $h$=2.8s, which is adequate but not comparably long. For $h$=2.8s, the MPC solution decelerates earlier with larger deceleration values compared to the benchmark IPO solution.

\begin{figure}[!htbp]
\centering
\includegraphics[width=3.4in]{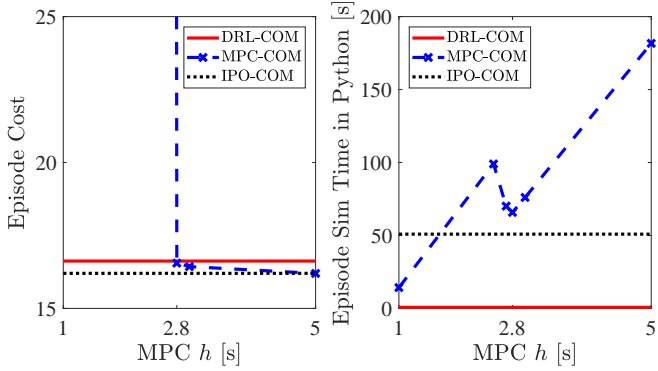}
\caption{Episode cost and simulation time in Python for the single initial condition considered.}
\label{fig:com_sum}
\end{figure}

\begin{table}[!htbp]
\caption{Comparison of DRL, MPC, and IPO solutions when there are no modeling errors for the single initial condition considered.}
\begin{center}
\begin{tabular}{|p{1.7cm}|p{3.6cm}|p{2.2cm}|}
\hline
 & Episode cost increase compared to the IPO benchmark [\%] & Episode simulation time in python [s]\\
\hline
DRL & 2.6 & 0.7\\
\hline
MPC ($h$=2.5s) & 1413.1  & 99.0\\
\hline
MPC ($h$=2.7s) & 1370.3  & 70.1\\
\hline
MPC ($h$=2.8s) & 2.2  & 65.8\\
\hline
MPC ($h$=3s) & 1.4 & 76.0\\
\hline
MPC ($h$=5s) & -0.02 & 181.7\\
\hline
IPO & 0 & 50.9\\
\hline
\end{tabular}
\end{center}
\label{table:com_compare}
\end{table}

The MPC episode cost decreases with the increasing prediction horizon, see Fig.~\ref{fig:com_sum} and Table~\ref{table:com_compare}. There is an acute change of the episode cost when increasing the prediction horizon from 2.7s to 2.8s, see Table~\ref{table:com_compare}. This may be because the defined cost function is almost the same as an absolute-value cost function, which generates acute MPC performance improvement around a certain prediction horizon. We experimented with a quadratic cost function whose results are not shown here; the transition of the episode cost with increasing prediction horizon values are smoother and slower.

From Table~\ref{table:com_compare}, the DRL episode cost is 2.6\% higher than that of the benchmark IPO solution, while MPC with the prediction horizon $h$=5s is equivalent to IPO ( 0.02\% lower). This suggests that DRL is equivalent to MPC with a sufficiently long prediction horizon with regards to the episode cost. From Table~\ref{table:com_compare}, we also see that the DRL-trained policy (which is a neural net) took 0.7 seconds in execution while the MPC and IPO optimization methods took at least 50 seconds, given that all testing was carried out in python on the same computer. Since setting the MPC prediction horizon $h$=5s produces nearly identical results as the benchmark IPO solution, we select $h$=5s for testing in the rest of the paper.

\subsection{Testing without modeling errors - different initial conditions}

Since a single initial condition may not be sufficient to demonstrate the average performance of DRL, different initial conditions are considered during testing, see Table~\ref{table:normal_cutin}. Two cases, the normal car-following and cut-in scenarios, are studied. For normal car following, the testing and training initial conditions have the same range. That is, -5$\leq e_0 \leq$5, -5$\leq e_{v,0} \leq$5, and -3$\leq a_{i,0} \leq$2 for testing normal car following. For cut-in scenarios, the negative initial gap-keeping error $e_0$ values are smaller than those of the training initial conditions while $e_{v,0}$, and $a_{i,0}$ have the same range as the training ones. In other words, -20$\leq e_0 \leq$-10, -5$\leq e_{v,0} \leq$5, and -3$\leq a_{i,0} \leq$2 for testing cut-in scenarios. For testing either the normal car-following or cut-in scenarios, we consider 5 values of $e_0$, 5 values of $e_{v,0}$, and 3 values of $a_{i,0}$, see Table~\ref{table:normal_cutin}. Thus, there are a total of 75 initial conditions, corresponding to 75 episodes.

The episode costs for either DRL, MPC or IPO are averaged over the 75 initial conditions to obtain the average episode cost. Both the DRL and MPC average episode costs are compared to that of the IPO. For normal car following, the DRL average episode cost is 5.8\% larger than the IPO benchmark, indicating near-optimal DRL-trained policy. For cut-in scenarios, the DRL average episode cost is 17.2\% larger than the IPO benchmark, indicating reduced optimality. This proves that the generalization issue of machine learning exists in the DRL-trained neural net. The MPC average episode cost is almost the same as the IPO benchmark with less than 0.1\% difference, indicating that the chosen prediction horizon $h$=5s is sufficiently long for all initial conditions.

\begin{table*}[!htbp]
\caption{COM testing without modeling errors for different initial conditions.}
\begin{center}
\begin{tabular}{|c|c|c|c|c|c|}
\hline
\multirow{2}{*}{Testing scenarios} &  \multicolumn{3}{|c|}{Initial conditions} & \multicolumn{2}{|c|}{Average episode cost increase [\%]} \\
\cline{2-6}
 & $e_0$ [m] & $e_{v,0}$ [m/s] & $a_{i,0}$ [m/s$^2$] & DRL compared to IPO & MPC ($h$=5s) compared to IPO \\
\hline
Normal car-following & \{-5,-2.5,0,2.5,5\} & \multirow{2}{*}{ \{-5,-2.5,0,2.5,5\} } & \multirow{2}{*}{ \{-3,0,2\} } & 5.8 & -0.1 \\
\cline{1-2} \cline{5-6}
Cut-in & \{-20,-17.5,-15,-12.5,-10\} &  &  & 17.2 & 0.0 \\
\hline
\multicolumn{6}{c}{Note: During training, $e_0$, $e_{v,0}$, and $a_{i,0}$ are randomly distributed in [-5,5]m, [-5,5]m/s, and [-3,2]m/s$^2$, respectively.}
\end{tabular}
\end{center}
\label{table:normal_cutin}
\end{table*}

\begin{figure}[!htbp]
\centering
\includegraphics[width=3in]{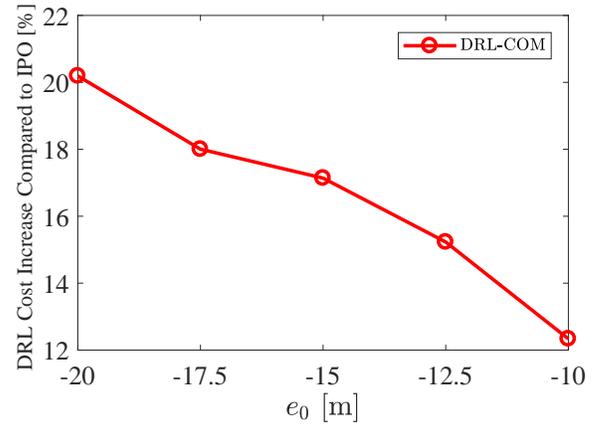}
\caption{DRL average episode cost increase compared to IPO for different initial gap-keeping error $e_0$ values in cut-in scenarios. Note that, during DRL training, -5$\leq e_0 \leq$5.}
\label{fig:cutin_sum}
\end{figure}

To further investigate the optimality reduction due to the difference between the testing and training initial conditions, we obtain the average episode cost increases with respect to different initial gap-keeping error $e_0$ values. In this case, for a certain $e_0$, the corresponding average episode cost is the average of the episode costs over different $e_{v,0}$ and $a_{i,0}$. The DRL and IPO average episode costs for each $e_0$ are compared, see Fig.~\ref{fig:cutin_sum}. It appears that the further-away $e_0$ is in testing compared to the $e_0$ range in training, the DRL average episode cost increase is higher as compared to the IPO benchmark, indicating more severe optimality reduction. This suggests that the DRL method is like curve-fitting based on available data; for testing outside the available data range, the fitted error grows as the fitted curve or the trained policy is not the true solution.

Note that MPC is model-based optimization and does not have the DRL generalization issue. For testing the initial conditions with larger magnitudes in cut-in scenarios, the average episode cost for MPC with the prediction horizon $h$=5s is almost the same (0.0\% difference) as the IPO benchmark, see Table~\ref{table:normal_cutin}.

\subsection{Testing with control delays - different initial conditions}

This controlled experiment studies the impact of different control delay $\tau_d$ values on the control performances of the DRL-trained policy and MPC. Compared to the ACC state-space equations, the only change is adding the control delay $\tau_d$ to the COM for testing. Recall that the control delay $\tau_d$ delays the execution of the control action $u_i$ by $\tau_d$. We consider three $\tau_d$ values: $\tau_d$=0.1s for pure electric vehicles; $\tau_d$=0.2s for combustion-engine or hybrid electric vehicles \cite{lidstrom2012modular,ploeg2011design}; and $\tau_d$=0.4s for combustion-engine trucks \cite{nieuwenhuijze2012cooperative}. For each $\tau_d$, the average episode cost is obtained over the 75 initial conditions.

\begin{figure}[!htbp]
\centering
\includegraphics[width=3in]{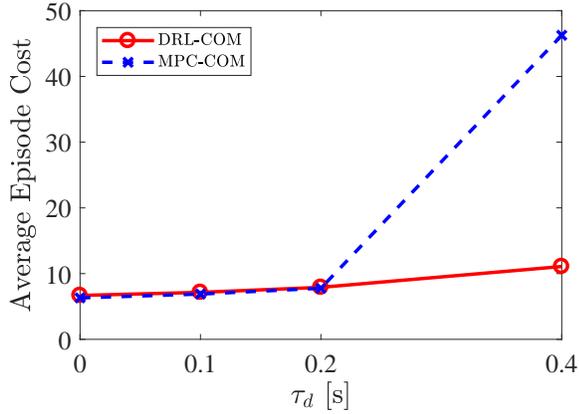}
\caption{Average episode cost for different control delay $\tau_d$ values. The control delay was added to the COM for testing.}
\label{fig:delay_sum}
\end{figure}

Fig.~\ref{fig:delay_sum} shows the DRL and MPC average episode costs for the increasing $\tau_d$ values. For smaller control delays $\tau_d\leq$0.2s, DRL and MPC have similar costs. For the largest delay $\tau_d$=0.4s, MPC has a significantly higher average episode cost. Through observing the time response of the states for $\tau_d$=0.4s, we found that the MPC control results have oscillatory steady state, leading to high costs. On the other hand, the DRL control results have zero steady-state errors, leading to low costs.

\subsection{Testing on the HFM - constant-speed following}

This controlled experiment studies the performance degradation of the COM-based DRL and MPC controllers when they are tested on the HFM, considering the differences between the COM and HFM. Compared to the ACC state-space equations, the only change is replacing the first-order COM with the HFM for testing. The first-order COM is an approximation to the acceleration command dynamics of the HFM. The modeling errors between the COM and the HFM of the 2015 Toyota Prius power-split PHEV have multiple causes. Firstly, the control delay of 0.2s is not considered in the COM. Secondly, the COM does not model mode switches between the electric and hybrid modes of Prius, which can happen during speeding and braking. Thirdly, the low-level PI controller causes overshoot and oscillation when correcting the error between the control input and actual acceleration, which is not modeled in the COM. Last but not least, the COM does not model the power limit of the vehicle, which could happen at high speeds. The power limit contributes to the largest modeling errors as it happens. Therefore, we consider HFM testing at different (low and high) speeds.

In the testing, the preceding vehicle has a constant speed with $a_{i-1}=0$. We consider the single initial condition as before: $[e_0, e_{v,0}, a_{i,0}]$=[5m,5m/s,0m/s$^2$]. During simulation, RK4 is used for HFM discretization. Note that the HFM operates at 100Hz, contrary to the update frequency of 10Hz ($\Delta t$=0.1s as the time step) of the COM.

\begin{figure}[!htbp]
\centering
\includegraphics[width=3.2in]{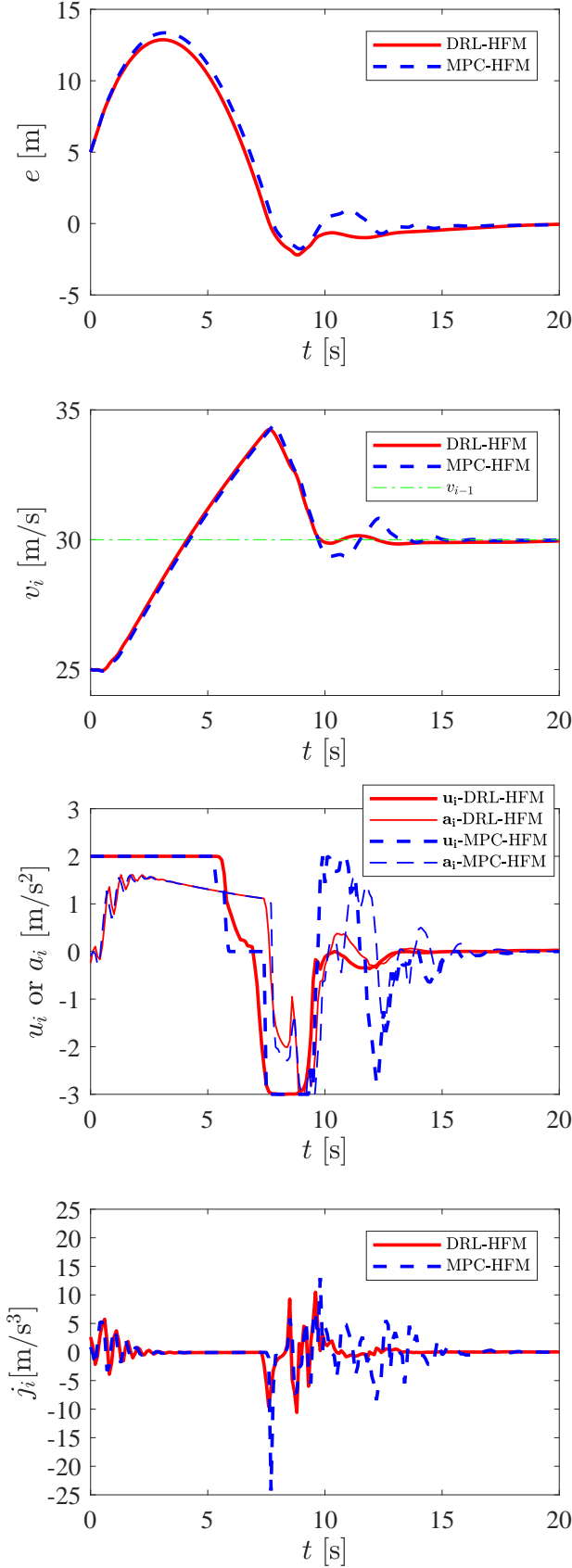}
\caption{HFM test results for constant-speed following for the single initial condition considered. The initial velocity of the following vehicle is $v_{i,0}$=25m/s. In the plots, MPC $h$=5s.}
\label{fig:hfm_episode}
\end{figure}

\begin{figure}[!htbp]
\centering
\includegraphics[width=3in]{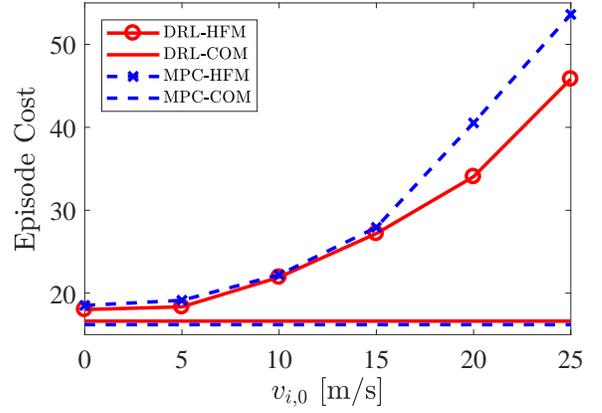}
\caption{Episode cost for constant-speed following for the single initial condition considered. In the plot, MPC $h$=5s.}
\label{fig:hfm_sum}
\end{figure}

The plots in Fig.~\ref{fig:hfm_episode} show the HFM testing results when the initial speed of the following vehicle $v_{i,0}$ is 25m/s. Correspondingly, the preceding vehicle is running at 30m/s. For $v_{i,0}\leq$15m/s, the DRL and MPC results are similar, see Fig.~\ref{fig:hfm_sum}. For $v_{i,0}$=25m/s, the following vehicle experienced the power limit constraint, see the third plot of control input and acceleration values in Fig.~\ref{fig:hfm_episode}. In this plot, the MPC control input exhibits a large jump in magnitude at 10s, while the DRL control input does not. This difference results in a significantly larger MPC episode cost, see Fig.~\ref{fig:hfm_sum}. Through another set of simulations, we found that the large jump of control input by MPC at $v_{i,0}$=25m/s could be eliminated by setting the control delay to zero, which is however impractical.

The plots in Fig.~\ref{fig:hfm_episode} also indicate the impact of modeling errors on the degradation of control performances. As the modeling errors increase with increasing speeds, both DRL and MPC episode costs increase, see Fig.~\ref{fig:hfm_sum}. However, for the same speed (same level of modeling errors), the MPC episode cost increases more than that of DRL, especially for high speeds. This suggests that DRL is better than MPC in coping with modeling errors in general. Note that, with the smallest speed $v_{i,0}$=0m/s, the DRL episode cost is higher than that from the COM testing. This is mainly due to the 0.2s control delay of the HFM that is not considered in the COM.

\subsection{Testing on the HFM - drive cycles}

This controlled experiment studies the performance degradation of the COM-based DRL and MPC controllers when they are tested on the HFM with drive cycles. For drive-cycle testing, the preceding vehicle's speed follows the drive cycle and is not constant. The non-zero $a_{i-1}$ results in more modeling errors, as we assume $a_{i-1}$=0m/s in the ACC state-space equations. The controllers were tested using three drive cycles, the HWFET, the FTP-75, and the US06. For the HWFET and FTP-75 drive cycles, the preceding vehicle's acceleration $a_{i-1}$ is within the range of the following vehicle's control input range [-3,2]m/s. For the US06 drive cycle of aggressive driving, the resulted $a_{i-1}$ exceeds the range of $u_i$ significantly. This also contributes to the modeling errors since the controllers are not able to generate comparable control input values for the following vehicle to match the acceleration of the preceding vehicle.

\begin{figure}[!htbp]
\centering
\includegraphics[width=3.2in]{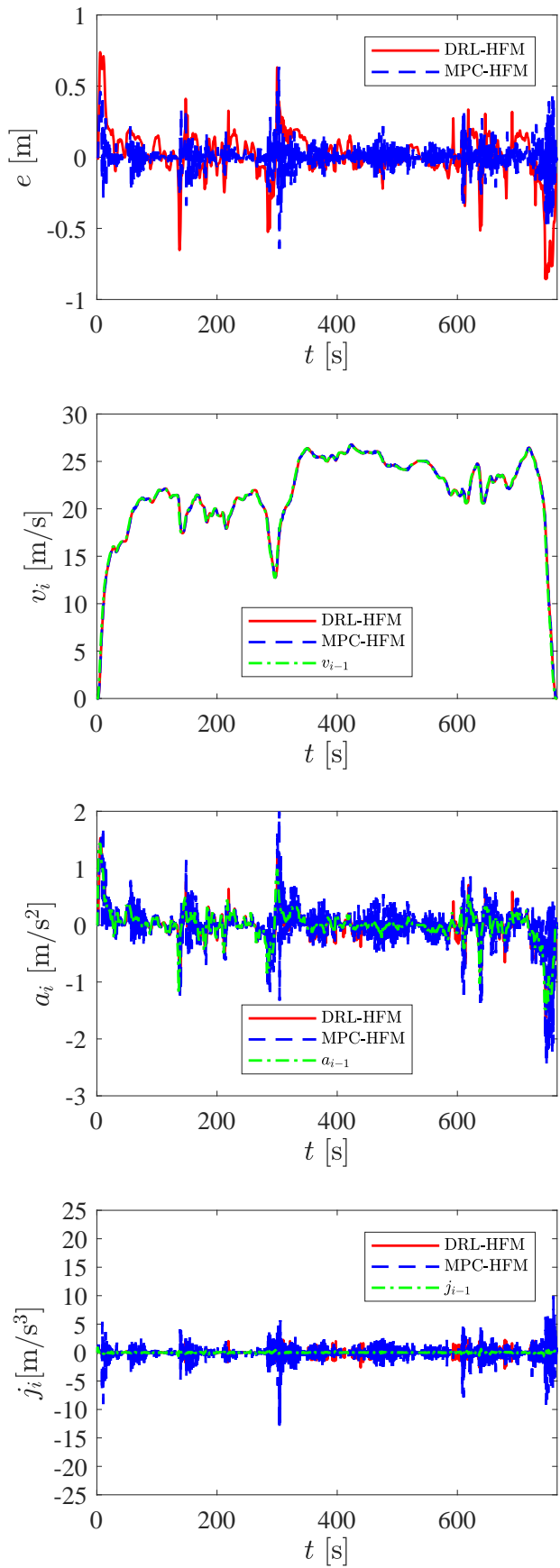}
\caption{HWFET drive-cycle test results on the HFM. In the plots, MPC $h$=5s.}
\label{fig:hwfet_episode}
\end{figure}

For drive-cycle testing, the initial condition is $[e_0, e_{v,0}, a_{i,0}]$ = [0m,0m/s,0m/s$^2$] for all drive cycles. Fig.~\ref{fig:hwfet_episode} shows the results of HFM testing with the HWFET drive cycle. The plots for FTP-75 and US06 drive cycles look similar, which are not shown here. From Table~\ref{table:drive_cycles}, for all three drive cycles, MPC have smaller gap-keeping errors but larger jerk magnitudes than DRL. Overall, the MPC episode costs for the three drive cycles are consistently larger than those of the DRL. The cost increase is 52.2\% for the HWFET highway-driving drive cycle while it is only 7.5\% for the FTP-75 city-driving drive cycle. The cost increase is smaller for FTP-75 because city driving results in smaller modeling errors due to less frequent power limit and control input saturation. The drive-cycle testing results support the suggestion that DRL is significantly better than MPC in coping with larger modeling errors.

\begin{table}[!htbp]
\caption{Drive-cycle test results on the HFM.}
\begin{center}
\begin{tabular}{|p{0.8cm}|p{0.8cm}|p{1.7cm}|p{1.8cm}|p{1.6cm}|}
\hline
Drive cycle & Method & $e_{min}, e_{mean}, e_{max}$ [m] & $j_{min}, j_{mean}, j_{max}$ [m/s$^2$] & MPC ($h$=5s) episode cost increase compared to DRL [\%]\\
\hline
\multirow{2}{*}{HWFET} & DRL & -0.9, 0.0 ,0.7 & -2.8, 0.0, 2.2 & - \\
\cline{2-5}
 & MPC & -0.6, 0.0, 0.6 & -13.7, 0.0, 9.9 & 52.2 \\
\hline
\multirow{2}{*}{FTP-75} & DRL & -1.3, 0.0, 7.4 & -6.8, 0.0, 13.7 & - \\
\cline{2-5}
 & MPC & -0.8, 0.0, 1.4 & -24.0, 0.0, 20.8 & 7.5 \\
\hline
\multirow{2}{*}{US06} & DRL & -2.2, 1.0, 27.2 & -22.6, 0.0, 11.1 & - \\
\cline{2-5}
 & MPC & -1.2, 0.7, 21.8 & -31.3, 0.0, 19.3 & 22.2 \\
\hline
\end{tabular}
\end{center}
\label{table:drive_cycles}
\end{table}

\section{Conclusion}

In this work, we compare DRL and MPC performances for an ACC car-following control problem with three state variables. The DRL training and MPC optimization are based on the same COM-dependent state-space equations and cost function. They also share the same discretization method and control update frequency. We also keep the testing conditions the same for DRL and MPC. The only difference is that DRL utilizes a reward discount factor $\gamma=0.99$ for convergence purpose. The discount factor is not used for MPC. We conduct testing using both the COM and HFM of the vehicle wherein the COM is a first-order approximation of the acceleration command dynamics of the HFM.

Our results show that, without modeling errors (testing based on the exact ACC state-space equations) and if the testing inputs are within the training data range, DRL is equivalent to MPC with a sufficiently long prediction horizon. Additionally, the DRL average episode cost averaged over 75 episodes is just 5.8\% higher than that of the IPO benchmark. It suggests that DRL can be used to train a policy that is very close to the optimal. However, if the testing inputs fall outside the training data range, there is significant degradation of the DRL control performance. That means, DRL suffers from the known generalization issue of machine learning.

With modeling errors (testing with control delays, on the HFM, and/or with drive cycles), DRL is significantly better than MPC regarding episode costs when the modeling errors are significant. When the modeling errors are small, the DRL and MPC performances are similar. In general, DRL is more tolerant (with smaller cost increase) than MPC in the presence of the same modeling errors. We summarize the general comparison of DRL and MPC for optimal control in Table~\ref{table:general_comparison}.

\begin{table}[!htbp]
\caption{General comparison of DRL and MPC for optimal control.}
\begin{center}
\begin{tabular}{|p{2.5cm}|p{2.3cm}|p{2.5cm}|}
\hline
Comparison criteria & DRL & MPC \\
\hline
Solution form & Multi-layer neural nets & Model-based optimization \\
\hline
Online computation time & Low & High, especially for high-degree non-linear systems \\
\hline
Optimality-seeking capability & Near-optimal & Almost optimal with long prediction horizons \\
\hline
Generalization issue of machine learning & Reduced optimality & N/A \\
\hline
Handling of hard constraints & Under development & Yes \\
\hline
Handling of modeling errors, control delays, and/or disturbances & Better for larger errors & Worse if not robust MPC \\
\hline
\end{tabular}
\end{center}
\label{table:general_comparison}
\end{table}

This work showed the optimality seeking performance and modeling-error tolerance of DRL through comparison with MPC. However, the underlying reason is yet to be analyzed. The challenge lies in the lack of theoretical analysis of the relationship between neural nets and optimal control, which could become significant future work. Here we present a qualitative explanation to the superior handling of modeling errors of DRL. In DDPG, the reinforcement learning environment state transition, i.e., transition from the current to the next state, is based on expectation of probabilities although the state-action mapping is deterministic. The probabilistic state transition allows for environment stochasticity that can be represented as modeling errors. The inherent consideration of environment stochasticity in the DDPG algorithm would thus contribute to the better modeling-error tolerance.

As mentioned in the Introduction, DRL has been shown to perform well in high-dimensional tasks. Thus, one of the next steps is to compare DRL and MPC with a dynamic system of higher degrees of freedom. We would also incorporate the state-of-the-art advancements in robust and stochastic MPC, and transfer and meta-learning for the comparison. We may also consider training the DRL policy with environmental noise since it has been shown to achieve better results in testing with environment stochasticity \cite{jang2019simulation}. The DRL's lack of theoretical assurance is in contrast with MPC's base theory on modeling and optimization, although MPC has the shortcoming of computing burden for online optimization. Thus, future work could involve combining DRL and MPC to take advantage of their best features and alleviate the shortcomings of both. \cite{williams2017information,kamthe2017data,nagabandi2018neural,gros2019data}.

\section*{Acknowledgment}

The authors would like to thank Toyota, Ontario Centres of Excellence, and the Natural Sciences and Engineering Research Council of Canada for the support of this work.

\ifCLASSOPTIONcaptionsoff
  \newpage
\fi



%\bibliographystyle{IEEEtran}

\bibliographystyle{IEEEtran}
\bibliography{references}

%\begin{thebibliography}{1}


%\end{thebibliography}


\begin{IEEEbiography}[{\includegraphics[width=1in,height=1.25in,clip,keepaspectratio]{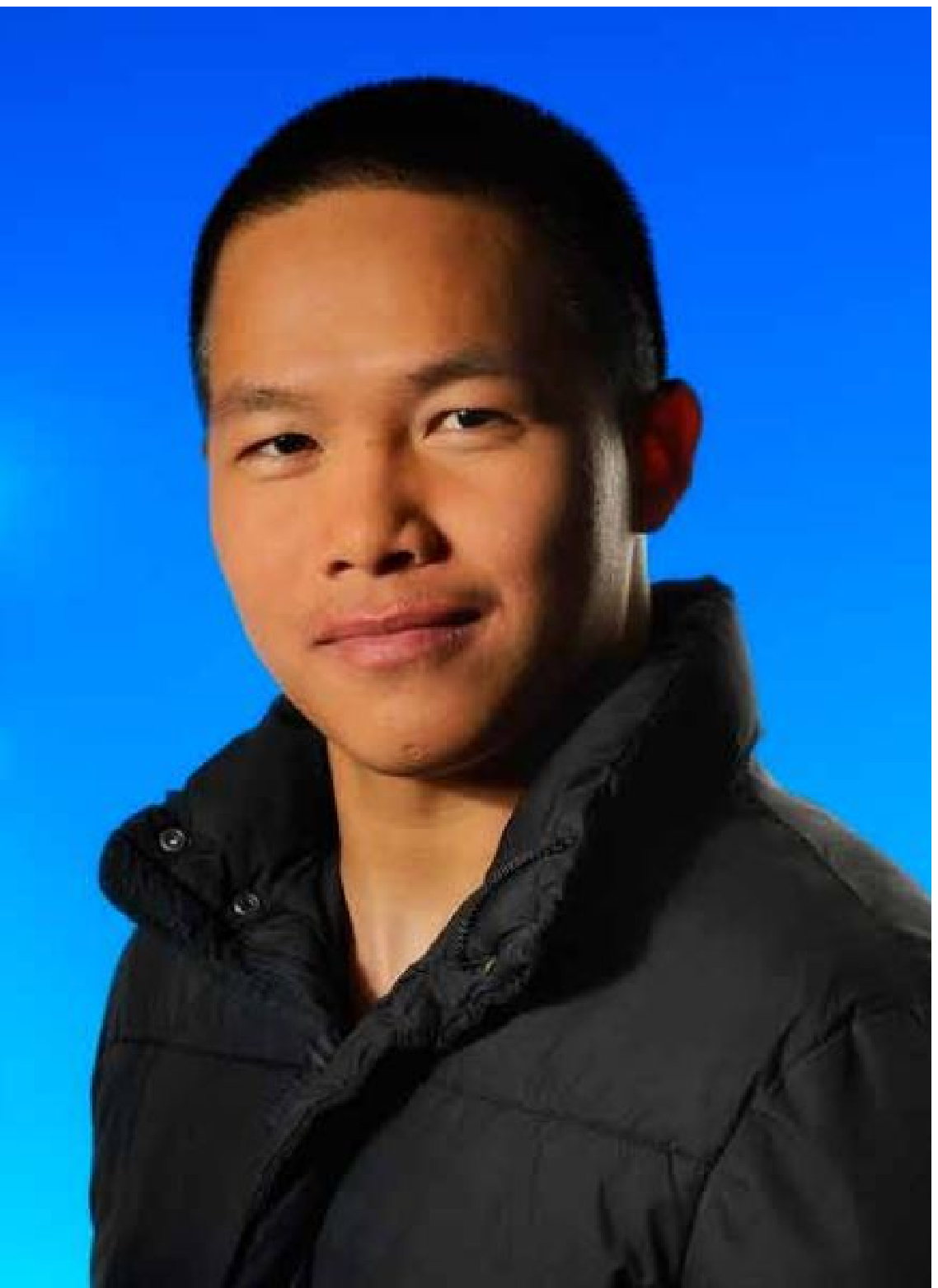}}]{Yuan Lin} received the B.E. degree in Civil Engineering from Nanchang University, China, in 2011 and the Ph.D. degree in Engineering Mechanics from Virginia Tech, Blacksburg, VA, USA, in 2016. He was a Postdoctoral Fellow in the Mechanical Engineering Department at Virginia Tech from 2016 to 2018. He is currently a Postdoctoral Fellow in the Systems Design Engineering Department at the University of Waterloo. His research interests include control, reinforcement learning, autonomous driving, and multi-agent systems. \end{IEEEbiography}

\begin{IEEEbiography}[{\includegraphics[width=1in,height=1.25in,clip,keepaspectratio]{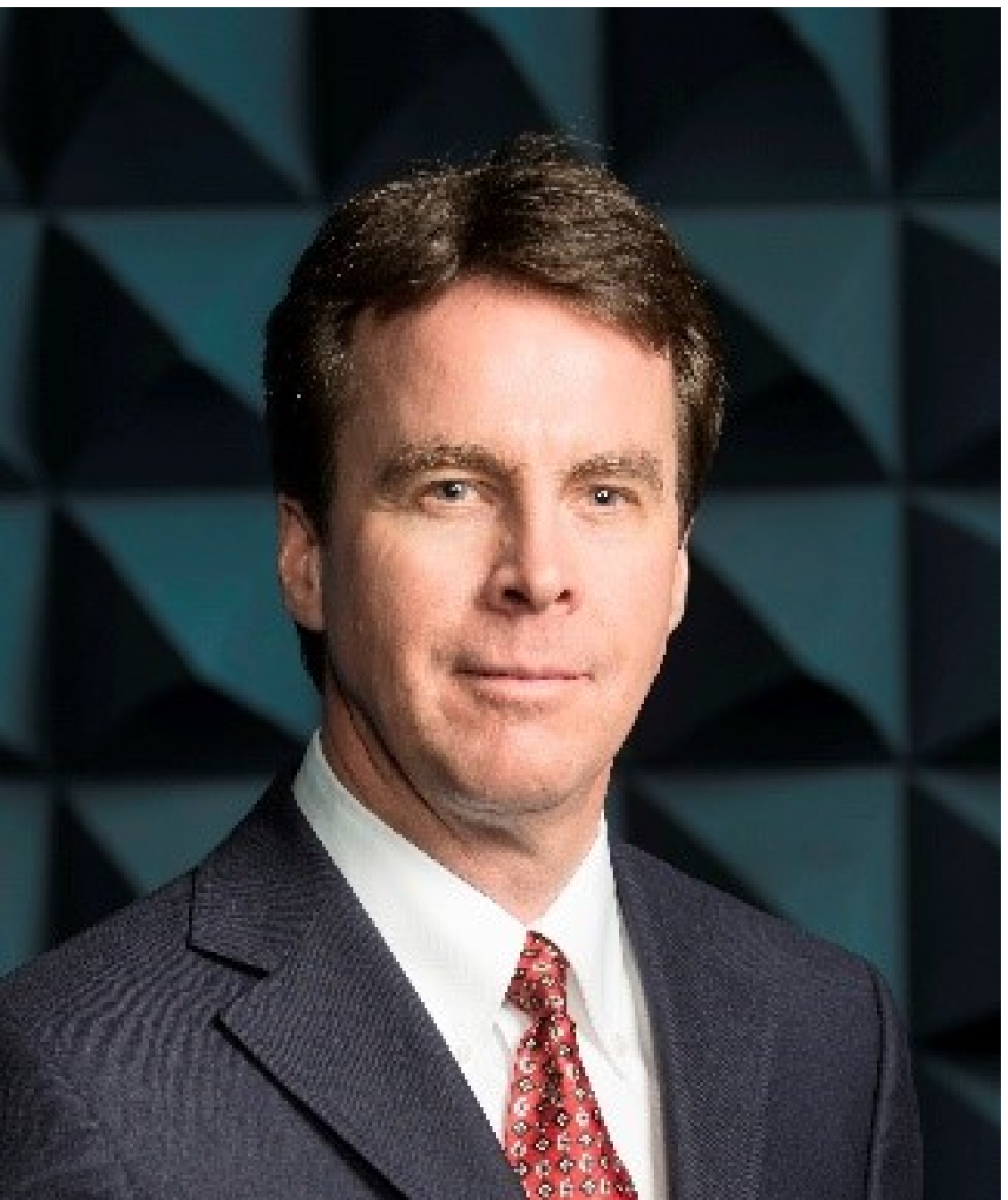}}]{John McPhee} is a Professor and the Canada Research Chair in System Dynamics at the University of Waterloo, Canada, which he joined in 1992. Prior to that, he held fellowships at Queen’s University, Canada, and the Université de Liège, Belgium. He pioneered the use of linear graph theory and symbolic computing to create real-time models and model-based controllers for multi-domain dynamic systems, with applications ranging from autonomous vehicles to rehabilitation robots and sports engineering. His research algorithms are a core component of the widely-used MapleSim modelling software, and his work appears in over 130 journal publications. Prof. McPhee is the past Chair of the International Association for Multibody System Dynamics, a co-founder of 2 international journals and 3 technical committees, a member of the Golf Digest Technical Panel, and an Associate Editor for 5 journals. He is a Fellow of the Canadian Academy of Engineering, the American and Canadian Societies of Mechanical Engineers, and the Engineering Institute of Canada. He has won 8 Best Paper Awards and, in 2014, he received the prestigious NSERC Synergy Award from the Governor-General of Canada. \end{IEEEbiography}

\begin{IEEEbiography}[{\includegraphics[width=1in,height=1.25in,clip,keepaspectratio]{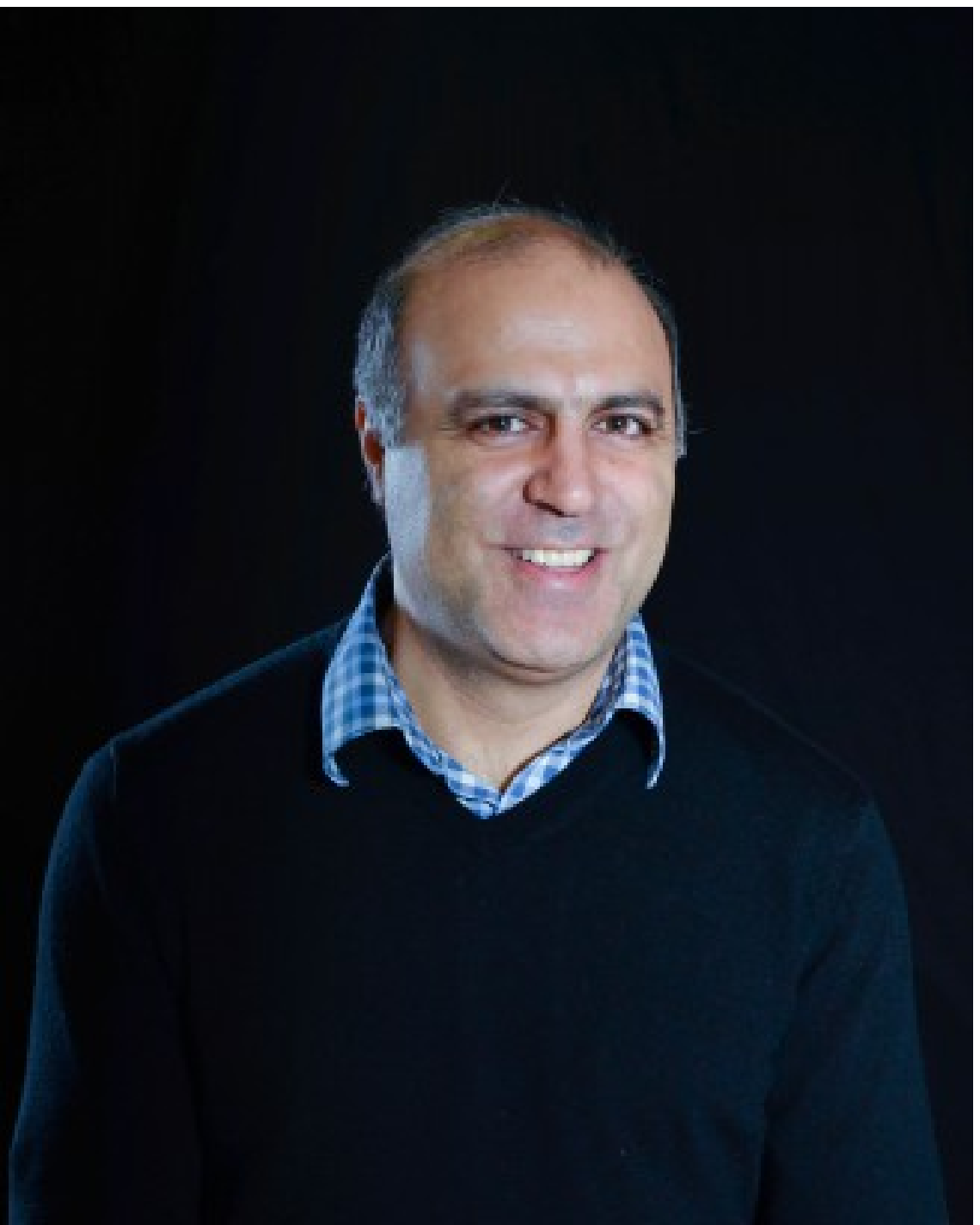}}]{Nasser L. Azad} is currently an Associate Professor in the Department of Systems Design Engineering, University of Waterloo, and the Director of Smart Hybrid and Electric Vehicle Systems (SHEVS) Laboratory. He was a Postdoctoral Fellow with the Vehicle Dynamics and Control Laboratory, Department of Mechanical Engineering, University of California, Berkeley, CA, USA. Dr. Azad’s primary research interests lie in control of connected hybrid and electric vehicles, autonomous cars, and unmanned aerial vehicle quad-rotors. He is also interested in applications of Artificial Intelligence for solving different engineering problems. Due to his outstanding work, Dr. Azad received an Early Researcher Award in 2015 from the Ministry of Research and Innovation, Ontario, Canada. \end{IEEEbiography}

%\begin{IEEEbiographynophoto}{Azim Eskandarian}
%Dr. Azim Eskandarian is a professor and the Department Head of Mechanical Engineering at Virginia Tech.
%% insert where needed to balance the two columns on the last page with
%%\newpage


%\vfill


\end{document}